\documentclass[twocolumn,pre,superscriptaddress]{revtex4}

\usepackage{graphics}
\usepackage{graphicx}
\usepackage{amsfonts}
\usepackage{textcomp}
\usepackage{amssymb}
\usepackage{mathrsfs}
\usepackage{amsmath}
\usepackage{color}
\usepackage{ulem}

\begin{document}

\title{Modeling cell migration regulated by cell-ECM micromechanical coupling}

\author{Yu Zheng\footnote{These authors contributed equally to this work.}}
\affiliation{Department of Physics, Arizona State University,
Tempe, AZ 85287}
\author{Hanqing Nan\footnotemark[1]}
\affiliation{Materials Science and Engineering, Arizona State
University, Tempe, AZ 85287}
\author{Qihui Fan\footnotemark[1]}
\affiliation{Beijing National Laboratory for Condensed Matte
Physics and CAS Key Laboratory of Soft Matter Physics, Institute
of Physics, Chinese Academy of Sciences, Beijing 100190, China}
\affiliation{School of Physical Sciences, University of Chinese
Academy of Sciences, Beijing 100049, China}
\author{Xiaochen Wang}
\affiliation{Beijing National Laboratory for Condensed Matte
Physics and CAS Key Laboratory of Soft Matter Physics, Institute
of Physics, Chinese Academy of Sciences, Beijing 100190,
China}\affiliation{School of Physical Sciences, University of
Chinese Academy of Sciences, Beijing 100049, China}
\author{Liyu Liu}
\affiliation{College of Physics, Chongqing University, Chongqing
401331, China}
\author{Ruchuan Liu}
\email[correspondence sent to: ]{phyliurc@cqu.edu.cn }
\affiliation{College of Physics, Chongqing University, Chongqing
401331, China}
\author{Fangfu Ye}
\email[correspondence sent to: ]{fye@iphy.ac.cn}
\affiliation{Beijing National Laboratory for Condensed Matte
Physics and CAS Key Laboratory of Soft Matter Physics, Institute
of Physics, Chinese Academy of Sciences, Beijing 100190, China}
\affiliation{School of Physical Sciences, University of Chinese
Academy of Sciences, Beijing 100049, China}
\author{Bo Sun}\email[correspondence sent to: ]{sunb@physics.oregonstate.edu }
\affiliation{Department of Physics, Oregon State University,
Corvallis, OR 97331}
\author{Yang Jiao} \email[correspondence sent to: ]{yang.jiao.2@asu.edu}
\affiliation{Materials Science and Engineering, Arizona State
University, Tempe, AZ 85287} \affiliation{Department of Physics,
Arizona State University, Tempe, AZ 85287}

\begin{abstract}

Cell migration in fibreous extracellular matrix (ECM) is crucial
to many physiological and pathological processes such as tissue
regeneration, immune response and cancer progression. During
migration, individual cells can generate active pulling forces via
actin filament contraction, which are transmitted to the ECM
fibers through focal adhesion complexes, remodel the ECM, and
eventually propagate to and can be sensed by other cells in the
system. The microstructure and physical properties of the ECM can
also significantly influence cell migration, e.g., via durotaxis
and contact guidance. Here, we develop a computational model for
cell migration regulated by cell-ECM micro-mechanical coupling.
Our model explicitly takes into account a variety of cellular
level processes including focal adhesion formation and
disassembly, active traction force generation and cell locomotion
due to actin filament contraction, transmission and propagation of
tensile forces in the ECM, as well as the resulting ECM
remodeling. We validate our model by accurately reproducing
single-cell dynamics of MCF-10A breast cancer cells migrating on
collagen gels and show that the durotaxis and contact guidance
effects naturally arise as a consequence of the cell-ECM
micro-mechanical interactions considered in the model. Moreover,
our model predicts strongly correlated multi-cellular migration
dynamics, which are resulted from the ECM-mediated mechanical
coupling among the migrating cell and are subsequently verified in
{\it in vitro} experiments using MCF-10A cells. Our computational
model provides a robust tool to investigate emergent collective
dynamics of multi-cellular systems in complex {\it in vivo}
micro-environment and can be utilized to design {\it in vitro}
micro-environments to guide collective behaviors and
self-organization of cells.

\end{abstract}



\maketitle

\section{Introduction}







Cell migration in fibreous extracellular matrix (ECM) is a complex
dynamic process involving a series of intra-cellular and
extra-cellular activities including the development of filopodia,
formation of focal adhesion sites, locomotion due to actin
filament contraction, and detachment of the rear end \cite{ref12,
ref13}. Collective cell migration in a complex micro-environment
is crucial to many physiological and pathological processes
including tissue regeneration, immune response and cancer
progression \cite{ref1, ref2, ref3, ref4}. Besides the
well-established chemotaxis \cite{ref14}, the microstructure and
physical properties of the ECM can also significantly influence
cell migration via durotaxis \cite{ref15, ref16, Brown2009},
haptotaxis \cite{ref17}, and contact guidance \cite{ref18, ref19,
Tranquillo1993}. For example, in durotaxis, a cell can sense and
respond to the rigidity gradient in the local micro-environment,
which in turn guides its migration \cite{Brown2009}.


A migrating cell also generates active pulling forces
\cite{ref20}, which are transmitted to the ECM fibers via focal
adhesion complexes \cite{ref21, ref22, ref23}. Such active forces
remodel the local ECM, e.g., by re-orienting the collagen fibers,
forming fiber bundles and increasing the local stiffness of ECM
\cite{ref24, ref25, ref26, ref27, ref28, ref29, shaohua2019}.
Recent studies have indicated that a delicate balance among the
magnitude of the pulling forces, the cell-ECM adhesion strength,
and the ECM rigidity is required to achieve an optimal mode of
single cell migration \cite{ref30}. In a multi-cell system, the
pulling forces generated by individual cells can give rise to a
dynamically evolving force network (carried by the ECM fibers) in
the system \cite{ref8, Frey07, ref5, ref6, ref7, ref8, ref9,
ref10, ref11}. In other words, the active pulling forces generated
by individual cells can propagate in the ECM and can be sensed by
distant cells. This ECM-mediated mechanical coupling among the
cells could further influence the migration of the individual
cells, which in turn alters the ECM structure and properties, and
thus the tensile force network. This feedback loop between the
force network and cell migration could lead to a rich spectrum of
collective migratory behaviors.

A variety of computational models have been developed to
investigate the migration dynamics of both single cell and
multi-cellular systems \cite{model01, model02, model03} as well as
various sub-cellular processes involved in cell migration
\cite{model04, model05, model06, model07, model08, model09}. For
example, a migrating cell can be modelled as an ``active
particle'' whose dynamics is mainly determined by an active
self-propelling force, a random drift and various effective
particle-particle and/or particle-environment interactions
\cite{model10, model11}. A wide spectrum of collective dynamics
have been observed and investigated in active-particle systems
\cite{model11}. On the other hand, vertex-based models
\cite{model12} and multi-state cellular Potts models
\cite{model13} are usually employed to investigate the collective
dynamics of densely packed sheets of cells, including the
spontaneous cell sorting driven by differential adhesion and the
epithelial to mesenchymal transition (EMT). Recently, cellular
automaton models which explicitly consider the migration of
invasive tumor cells following least-resistance paths have been
devised to study the emergence of invasive dendritic structures
composed of highly malignant tumor cells emanating from the
primary tumor mass \cite{model14, model15, model16, model17}.

In the preponderance of existing cell migration models, the
influence of the cell-ECM interactions and/or ECM-mediated
indirect cell-cell interactions on collective migration dynamics
either is not considered or is incorporated in an effective
phenomenological manner. Recently, a computational model based on
continuum mechanics has been developed that explicitly considers
the micro-mechanical coupling of a migrating cell and the 2D
substrate \cite{model18}. Durotaxis effects have been successfully
reproduced from this model. Moreover, a novel model for
investigating cell migration in model 2D ECM network guided by
mechanical cues has been developed by considering coarse-grained
cytoskeleton of a migrating cell as a part of the ECM network and
the cells can hop between neighboring nodes of the network
\cite{ref31}. Even in these novel models which explicitly take
into account cell-ECM micro-mechanical couplings, a number of
processes crucial to cell migratory behaviors such as focal
adhesion formation and disassembly, actin filament contraction and
the resulting continuous cell locomotion, the remodeling of ECM
network and the influence of the complex microstructure and
topology of ECM network have not been explicitly considered and
incorporated into the models.

Here, we develop a computational model for cell migration
regulated by cell-ECM micro-mechanical coupling, which could be
employed to investigate collective migratory behaviors and
emergent self-organizing multi-cellular patterns resulted from
ECM-mediated mechanical signaling among the cells. Our model takes
into account a variety of cellular level processes including focal
adhesion formation and disassembly, active traction force
generation and cell locomotion due to actin filament contraction,
transmission and propagation of tensile forces in the ECM. We
employ a node-bond (i.e., graph) representation to model the
complex 3D ECM network microstructure, which is reconstructed
based on confocal imaging data. In addition, we use a nonlinear
mechanical model for the ECM networks, which incorporates buckling
of collagen fibers upon compression and strain-hardening upon
stretching. We validate our model by accurately reproducing
single-cell dynamics of MCF-10A breast cancer cells migrating on
collagen gels and show that the durotaxis and contact guidance
effects naturally arise as a consequence of the cell-ECM
micro-mechanical interactions considered in the model. Moreover,
our model predicts strongly correlated multi-cellular migration
dynamics, which are resulted from the ECM-mediated mechanical
coupling among the migrating cell and are subsequently verified in
{\it in vitro} experiments using MCF-10A cells.


The rest of the paper is organized as follows: In Sec. II, we
describe the microstructural and mechanical model of the 3D ECM
(mainly collagen I) networks. In Sec. III, we introduce our cell
migration model and discuss the associated assumptions and
limitations. In Sec. IV, we validate our model by producing
single-cell migration dynamics of MCF-10A breast cancer cells on
isotropic collagen network and investigate the cell migration
dynamics on heterogeneous networks with stiffness gradient and
aligned fibers. In Sec. V, we investigate collective
multi-cellular dynamics resulted from ECM-mediated mechanical
coupling among the migrating cells, and validate our results via
{\it in vitro} experiments. In Sec. VI, we provide concluding
remarks.








\section{Microstructure and mechanical model of 3D ECM network}

\subsection{Modeling ECM Network via Statistical Descriptors and Stochastic Reconstruction}

In this section, we briefly describe the microstructural and
micro-mechanical models for the ECM networks. The detailed
descriptions of these models are provided in Refs. \cite{ref33}
and \cite{ref11}, respectively. The 3D ECM, mainly composed of
collagen type I gel, is modeled as a discrete network with a
``graph'' (i.e., node-bond) representation in a cubic simulation
domain with linear size $L$ ($\sim 300 \mu m$), which is composed
of $M_n$ nodes and $M_b$ bonds, depending on the collagen
concentration. The average coordination number $Z$, i.e., the
average number of bonds connected to each node, is given by $Z =
2M_b/M_n$. We mainly use fixed boundary (FB) conditions (i.e., the
nodes within a certain distance $\delta L \sim 5 \mu m$ from the
boundaries of the simulation domain are fixed) in our simulations,
but also confirm that using periodic boundary (PB) conditions does
not affect the results for the large $L$ values used in our
simulations.

We employ a set of statistical descriptors for quantifying the
network geometry and topology \cite{ref33}, which include the node
density $\rho$ (corresponding to the collagen concentration), the
fiber (or bond) length distribution function $P_f$, the
distribution of coordination number (i.e., the number of neighbors
of a node) $P_Z$, and the average fiber orientation $\Omega$
(measured as the average cosine value associated with the acute
angle of a fiber made with respect to a prescribed direction).
These statistical descriptors can be computed from 3D ECM network
extracted from confocal images via skeletonization techniques
\cite{ref25}. Fig. 1(c) and (d) respectively shows the
coordination distribution $P_Z$ and fiber length distribution
$P_f$ for homogeneous collagen networks with a collagen
concentration of $2 mg/ml$, which will be used in our subsequent
investigations. The average fiber length is $1.96 \mu m$ and the
average coordination number $Z = 3.4$. Since the fibers are
randomly oriented in homogeneous networks, the average fiber
orientation metric $\Omega \approx 0.5$. The node number density
$\rho \approx 0.185 /\mu m^3$.

\begin{figure}[ht]
\includegraphics[width=0.495\textwidth,keepaspectratio]{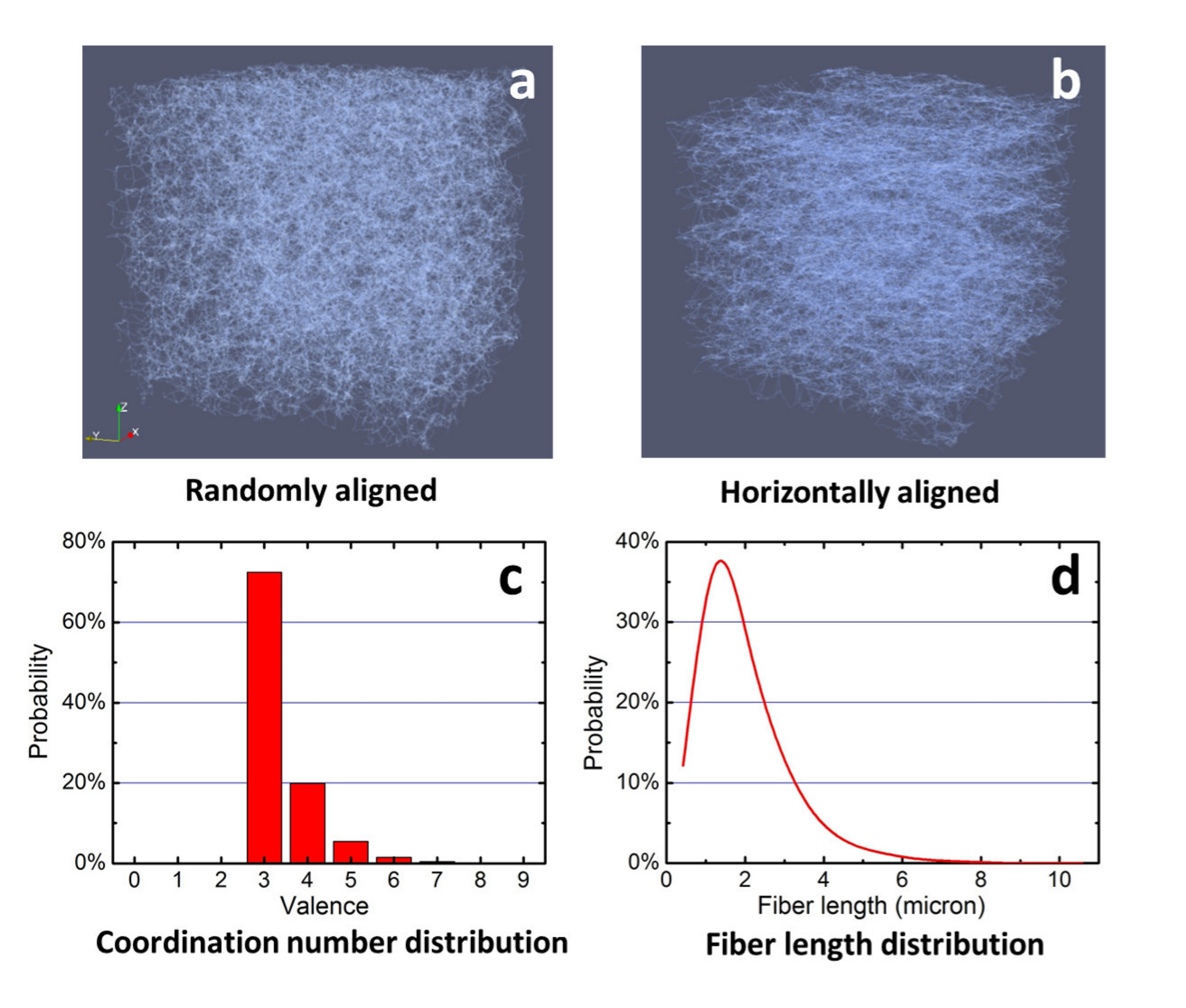}
\label{fig_1}\caption{Realizations of 3D ECM networks with
randomly oriented fibers with $\Omega = 0.5$ (a) and horizontally
aligned fibers with $\Omega = 0.88$ (b), generated using
stochastic reconstruction techniques. For better visualization,
only small sub-networks with a linear size of $50 \mu m$ are shown
here. (c) and (d) respectively shows the coordination distribution
$P_Z$ and fiber length distribution $P_f$ for the networks, which
are computed based on confocal images of a 2 mg/ml collagen gel.}
\end{figure}

For a given set of network statistics (e.g., $P^*_Z$, $P^*_f$,
$\Omega^*$ and $\rho^*$), we can generate realizations of the
networks associated with the prescribed descriptors using
stochastic reconstruction \cite{ref33}. In particular, we start
from a randomly generated initial network with the prescribed node
number density $\rho^*$. From this initial network, the
descriptors $P_f$, $P_Z$, and $\Omega$ are computed and compared
to the corresponding prescribed descriptors. An energy functional
$E$ is defined as the sum of the squared differences between the
computed and corresponding prescribed the descriptors
\cite{ref33}, i.e.,
\begin{equation}
E = \sum_r |P_Z(r) - P^*_Z(r)|^2 + \sum_r |P_f(r) - P^*_f(r)|^2 +
|\Omega - \Omega^*|^2.
\end{equation}
Next, the initial network is perturbed by randomly displacing a
node and/or removing/adding a bond to randomly selected pairs of
nodes. A new energy for the new network is computed. If the new
energy $E_{new}$ is lower than the old energy $E_{old}$, the new
network replaces the old one. Otherwise, the new network
configuration replaces the old network with the probability
$e^{(E_{old}-E_{new})/T}$, where $T$ is a virtual temperature,
which possesses an initial large value and is gradually decreased.
The network is continuously evolved in this way (more precisely,
via simulated annealing method \cite{sim_annealing} to allow even
energy-increasing network during the initial stages) until $E
\approx 0$, i.e., the computed descriptors match t he prescribed
ones within a prescribed small tolerance. The detailed of this
technique is provided in Ref. \cite{ref33}.

We note that one can either use experimentally obtained network
statistics as the target descriptors in the reconstruction or can
construct a set of feasible hypothetical statistical descriptors
in order to control the geometry and topology of the constructed
random network. Fig. 1(a) shows a reconstructed network based on
the experimentally obtained statistics of the $2 mg/ml$ collagen
gels, in which the fibers are randomly oriented. In order to
investigate the effects of fiber alignment on cell migration
dynamics, we also generate realizations of networks with
horizontally aligned fibers (see Fig. 1(b)). This is achieved by
setting $\Omega^* = 1$ with respect to the x-direction, and using
the same $P^*_Z$, $P^*_f$ and $\rho^*$ of the homogeneous network.
We note that the optimized $\Omega$ of the reconstructed network
is in fact a little smaller than unity ($\Omega \sim 0.88$), due
to additional topological and geometrical constraints specified by
$P^*_Z$ and $P^*_f$. Nevertheless, the fiber alignment is already
very significant in the reconstructed networks.

\subsection{Micro-mechanical Model of ECM Networks}

The ECM (collagen) fibers are highly non-linear, typically
exhibiting buckling, strain-hardening and plastic behaviors
\cite{ref25, MacKintosh05, Safran12, nat_method15}, which can
significantly affect the propagation of the active forces in the
system. The nonlinearity of the ECM fibers also induces a
nontrivial coupling with the cell contractility, i.e., for small
contraction, the fibers may be in the linear elastic regime, while
for large contraction, the fibers may be in the strain-hardening
or plastic regime \cite{ref11}. This in turns can affect the
overall cell migration dynamics \cite{ref30}.

\begin{figure}[ht]
\includegraphics[width=0.44\textwidth,keepaspectratio]{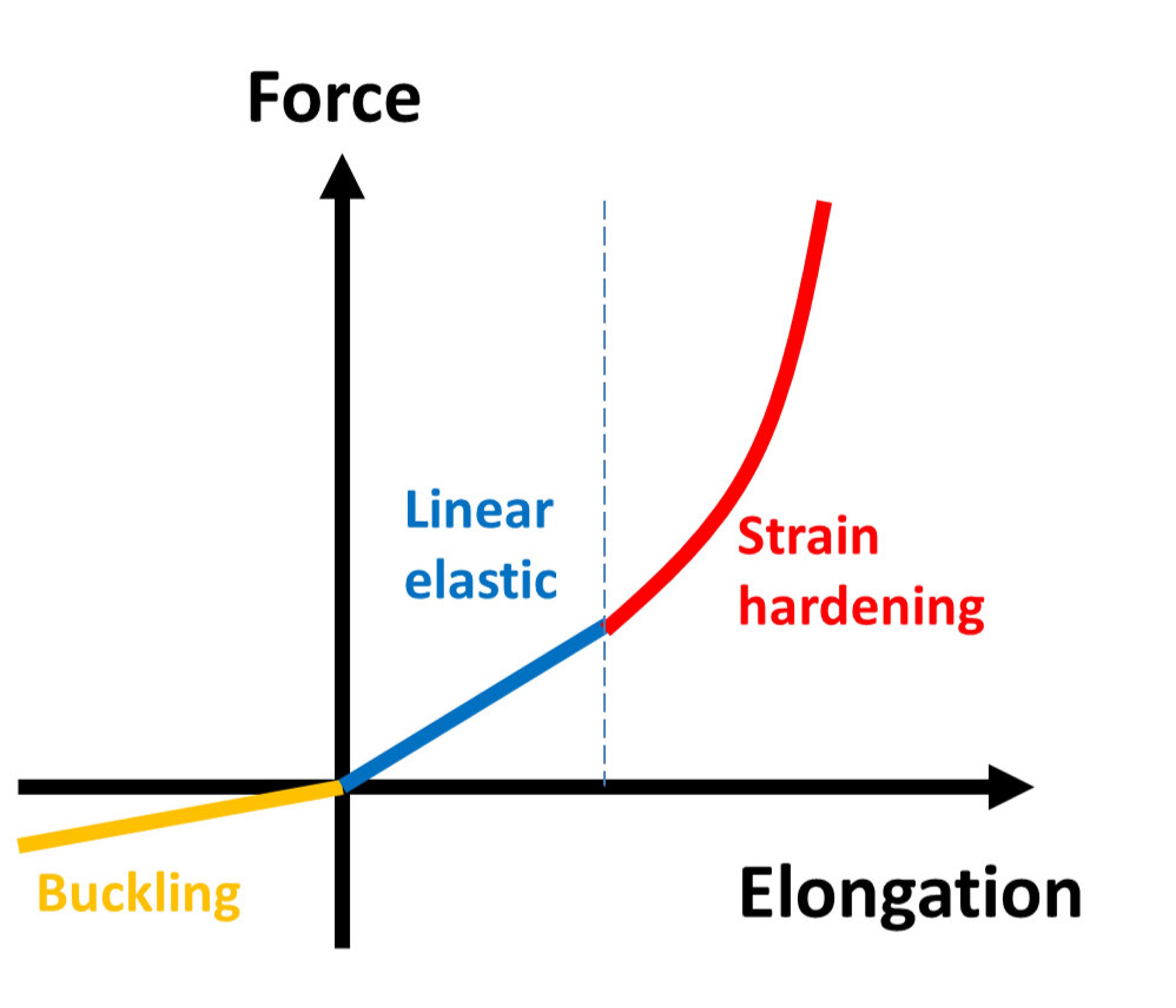}
\label{fig_2}\caption{Schematic illustration of the nonlinear
elastic fiber model, including a linear elastic regime, which is
followed by a strong strain-hardening regime upon compression and
buckling upon compression.}
\end{figure}

In this work, we will use a nonlinear micromechanical model for
the ECM fiber, which is schematically illustrated in Fig. 2
\cite{ref11, nat_method15}. In particular, upon stretching, a
fiber first enters a linear elastic regime, which is followed by a
strong strain-hardening regime once the elongation is larger than
a prescribed threshold. Upon compression, we consider the fiber
immediately buckles and thus, possesses a much smaller compression
modulus. The elongation stiffness $k$ of the fiber is thus given
by
\begin{equation}
\label{eq_nonlinear} k = \left\{{\begin{array}{cc} \rho EA, &
\lambda < 0
\\\\ EA, & 0<\lambda < \lambda_{s} \\\\ EA \exp[(\lambda
-   \lambda_{s})/\lambda_0], & \lambda > \lambda_{s}.
\end{array}}\right.
\end{equation}
where $E$ and $A$ are respectively the Young's modulus and
cross-sectional area of the fiber bundle, and we use $EA = 8\times
10^{-7} N$ \cite{ref25}; $\lambda = \delta \ell/\ell$ is
elongation strain, and $\lambda_s = 0.02$ and $\lambda_0 = 0.05$
are parameters for the strain-hardening model \cite{nat_method15};
$\rho = 0.1$ describes the effects of buckling \cite{ref11}. In
addition, we consider the fiber segments as well as the cross
links (nodes) can resist bending and employ a first-order bending
approximation \cite{Frey07}, for which the bending energy $E_b$ is
a function of transverse displacement $u$ of the two nodes of a
fiber, i.e., $E_b = \alpha EI u^2/\ell_0$, where the bending
modulus $EI = 5 \times 10^{-22} Nm^2$ \cite{ref25}, $I$ is the
second moment of area, $\ell_0$ is the original length of the
fiber segment, and $\alpha = 1.8$. We also note that the effects
of interstitial fluid, which quickly dissipates the kinetic energy
generated due to cell contraction, are not explicitly considered.


Plasticity of the fibers is modeled as a time-dependent elongation
of the fiber with a constant flow rate $\gamma$, i.e., $\delta l_P
= \gamma t$, once the stretching force on the fiber is larger than
a prescribed threshold $f_P$. The flow rate can be calibrated
based on experimental data available in literature \cite{ref29}.
We note that this elongation due to plasticity effectively reduces
the stiffness of the fiber, i.e., $k = EA/(l+\delta l_P)$. In
addition, we can easily construct a stiffness gradient in the ECM
network, by introducing a position-dependent scaling factor, i.e.,
$E(x) = E\cdot C_0(x)$, where $C_0(x) = (1 + x/L)$. It is clear
that other forms of $C_0(x)$ than the simple linear scaling could
be employed to model more stiffness gradient. In the subsequent
studies, we will use the simple constant gradient to investigate
the durotaxis effects.

Once the cell contractions are applied (as described in Sec. III),
an iteration procedure \cite{ref11} will be employed to find the
force-balanced state of the network and obtain the forces on the
fibers. The numerical procedure can be easily parallelized using
OpenMP for large networks.


\section{Modeling cell migration regulated by cell-ECM micro-mechanical coupling}




In this section, we present in detail the cell migration model,
which is coupled with the ECM network model. We note that the
current model is targeted for highly motile non-invasive cancer
cells, such as the MCF-10A breast cancer cells, moving on 3D
collagen gel (see Fig. 3 for illustration). In this case, the
migrating cells are strongly coupled with the ECM via their
micro-mechanical interactions without any ECM degradation, which
is very challenging to accurately model. We will briefly discuss
the generalization of the current model to incorporate ECM
degradation in Sec. VI.

\begin{figure*}[ht]
\centering
\includegraphics[width=0.95\textwidth]{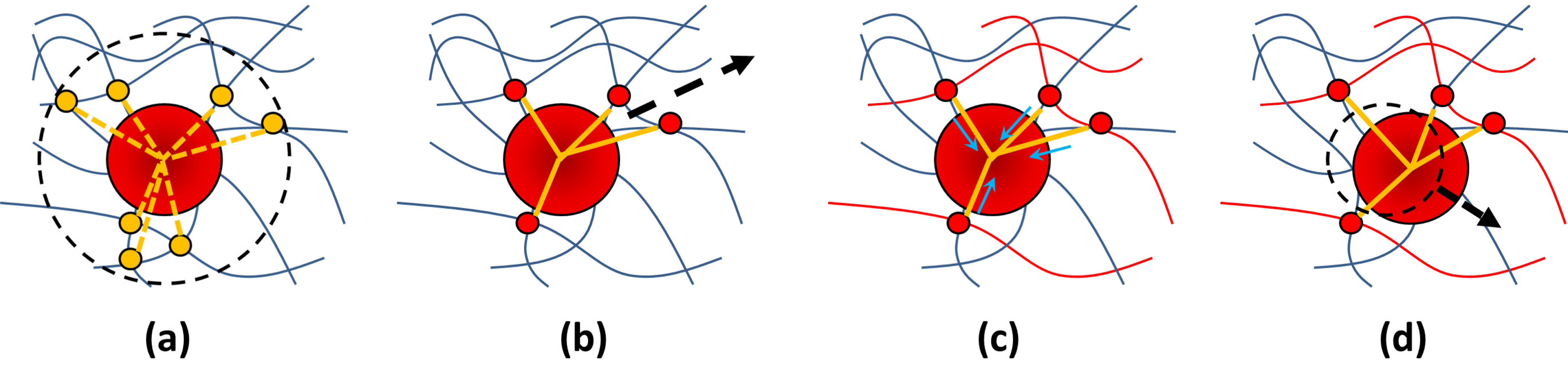}
\caption{Schematic illustration of the computational model for a
migrating cell. (a) Protrusions (yellow dashed lines) generated by
polymerization of actin filaments can lead to focal adhesion
formation (yellow dots) in a region within $\delta R_s$ (effective
protrusion length) from the cell surface (region enclosed by
dashed circle). (b) New focal adhesion (red dots) formation is
modeled by adding new filaments (yellow solid lines) connecting
the center of mass of the cell (i.e., the center of the cytoplasm
sphere) to a randomly selected node of the ECM network within
$\delta R_s$ from the cell surface, with a probability depending
on the persistent direction of the cell (dashed arrow) and the
local stress state of the fibers. (c) Active contraction of actin
filaments generates active forces in the ECM network and leads to
deformation of ECM. (d) Locomotion of the cell due to actin
filament contraction.} \label{cell_model}
\end{figure*}


As illustrated in Fig. 3, our cell model consists of an elastic
sphere representing the exclusion volume associated with cytoplasm
and a set of cytoskeleton filaments connecting the cytoplasm
sphere to the plasma membrane. As a starting point, we will not
distinguish various types of actin assemblies and microtubules for
simplicity, and only consider the contractility of the filaments.
The plasma membrane is then modeled as the minimal hull associated
with the end points of the filaments (see Fig. 3(A)). In the
beginning of the simulation, a cell (i.e., a cytoplasm sphere and
the associated cytoskeleton filaments) is introduced in the
collagen network, and a random persistent direction ${\bf n}_0$ is
selected. The migration process is decomposed into cycles of
successive events including (i) development of protrusion (due to
active actin filaments polymerization) and formation of new
adhesion sites, (ii) contraction of actin filaments and the
resulting locomotion of the cell, and (iii) breaking of old
adhesion sites. These events are modeled and simulated as
described below:



\indent$\bullet$ (i) Protrusions are generated by the elongation
(polymerization) of the actin filaments, which can be attached to
the ECM fibers via focal adhesion. This process is modeled by
adding new filaments connecting the center of mass of the cell
(i.e., the center of the cytoplasm sphere) to a randomly selected
node of the ECM network within $\delta R_s \sim 5 \mu m$
(effective protrusion length) from the cell surface (see Fig.
\ref{cell_model}(a) and (b)), with the probability $p_a$ given by
\begin{equation}
p_a = c_1 ({\bf n}_0\cdot{\bf d}) + c_2 \sigma_f
\end{equation}
where ${\bf n}_0$ is the persistent direction of the cell, ${\bf
d}$ is vector connecting the cell center and the network node,
$\sigma_f$ is the largest stress on the fibers connected to the
node, $c_1$ and $c_2$ are proportionality constants. This model
implies that actin polymerization is more likely to occur in the
polarized region of the cell \cite{Zallen2009}; and that it is
more likely to form an adhesion site on highly stressed fibers
\cite{Munro2011, Matsumura2004}. Each adhesion site has a finite
life span $T_a$ and breaks once $T_a$ is reached.


\indent$\bullet$ (ii) The contraction of an actin filament
connecting the cytoplasm sphere and ECM network can generate a
traction force ${\bf f}_t$ ($\sim 1 nN$) along the filament
\cite{boey1998, boey1998II, coughlin03, gordon2012} and a
shrinkage of the filament length $\delta l$ ($\sim 10\%$ of the
original length); see Fig. \ref{cell_model}(c). This active force
is transmitted to the ECM network through the ``focal adhesion''
node. Force boundary condition is imposed to this node (and other
nodes connected to contracting filaments) and the deformed
force-balanced network configuration is obtained as described in
Sec. II.B. The length $d'$ of the filament connecting the center
cell and the displaced node is then computed. We then consider the
contraction of this filament generates a displacement component
for the cell enter, i.e.,
\begin{equation}
\label{eq_fila} \delta {\bf x} = \max\{\delta l - (d-d'),
0\}\cdot{\bf d}'_0
\end{equation}
where $\delta l$ is intrinsic contraction of the filament, $d$ and
$d'$ are respectively the distance between the cell center and the
adhesion node before and after ECM deformation due to filament
contraction, and ${\bf d}'_0$ is the unitary direction vector
along the filament direction after ECM deformation.

\indent$\bullet$ (iii) Once the displacement components associated
with all filaments are computed, the center of mass position of
the cell is updated as follows (see Fig. \ref{cell_model}(d)):
\begin{equation}
\label{eq_locomotion} {\bf x}_{t+1} = {\bf x}_t + \sum_i \delta
{\bf x}_i
\end{equation}
where the sum is taken for all filaments, and $\delta {\bf x}_i$
is the displacement component associated with the $i$th filament.
The persistent direction ${\bf n}_0$ is updated as the direction
of the cell displacement (i.e., $\sum_i \delta {\bf x}_i$). We
note that Eq. (\ref{eq_fila}) and Eq. (\ref{eq_locomotion}) imply
that cell locomotion is due to actin filament contraction and
depends on the stiffness of the local ECM.






\indent$\bullet$  (iv) All of the current adhesion sites are
checked and those reach their life span $T_a$ are consider to
break, leading to the detachment of the cell surface from the
collagen fibers.


In the simulation, time is discretized such that a migration cycle
is completed during the elapsing of one time step $dt$. The life
time of focal adhesion sites $T_a = 2dt$, which is calibrated
based on the experimental data (see Sec. IV for details). Once an
entire migration cycle is completed, the position of cytoplasm
sphere (and thus, the center of mass of the cell) is updated and
the cell starts the next migration cycle, by repeating the steps
(i) to (iv).

We also note that the cell-cell contact adhesion is not explicitly
considered in this model, since our focus here is highly motile
breast cancer cells with very weak cell-cell adhesion. In
addition, we employ a minimal model for the contact inhibition
effect for multi-cellular systems. In particular, we consider that
if a pair of cells with radius $R_s$ ($\sim 10 \mu m$) overlap,
they feel a mutual repulsive force proportional to the linear
overlap distance, i.e., $F_r = \kappa \delta R$, where $\kappa$ is
an effective elastic constant depending on the modulus of the
cell, $\delta R = 2R_s - d_s$ is the overlap distance, and $d_s$
is the cell center separation distance. In the subsequent
sections, we will validate our model using single-cell migration
experiments and employ the model to predict multi-cell migration
dynamics.



\section{Single-cell migration dynamics}



In this section, we employ our model to investigate single cell
migration dynamics and its regulation by the microstructure and
mechanical properties of the micro-environment (i.e., the ECM
network). We mainly focus on MCF-10A breast cancer cells in our
simulations. The MCF-10A cell are highly motile non-metastatic
cancer cells which exhibit strong micro-mechanical coupling of ECM
networks when moving on collagen gels and do not degrade the
collagen fibers \cite{qihui2019}. Therefore, this system provides
an ideal system for testing our model. In the following
discussions, we will directly use the experimental results to
validate our model predictions. The experimental details are
provided in Ref. \cite{qihui2019}.

\subsection{Migration dynamics of MCF-10A cells on isotropic collagen gel}

We first employ our model to study the migration dynamics of
individual MCF-10A breast cancer cells on isotropic collagen gels
with randomly oriented fibers. It is well established that in the
case, the overall cell dynamics can be captured by the
active-particle model \cite{RMP}, i.e.,
\begin{equation}
\gamma d{\bf r}/dt = F\hat{{\bf e}} + \xi
\end{equation}
where ${\bf r}$ is the particle center of mass, $\gamma$ is an
effective friction coefficient, $F$ is an effective constant
self-propelling force, $\hat{\bf e}$ is the persistent direction
which is subject to a random rotational diffusion and $\xi$ is a
white-noise random vector \cite{RMP}. The associated theoretical
mean square displacement (MSD) $\sigma^2$ is given by \cite{RMP}
\begin{equation}
\label{eq_MSD} \sigma^2(t) =
[4D+2v^2\tau_R]t+2v^2\tau^2_R[e^{-t/\tau_R}-1]
\end{equation}
where $D$ is the diffusivity of the particle, $v$ is the
persistent velocity and $\tau_R$ is the relaxation time for
rotation diffusion of the persistent direction. It can be seen
from Eq. (\ref{eq_MSD}) that for small $t$, the particle exhibit
ballistic dynamics with $\sigma^2 \propto t^2$. At large $t$, the
system is diffusive, with $\sigma^2 \propto t$.

\begin{figure}[htp]
\includegraphics[width=0.48\textwidth]{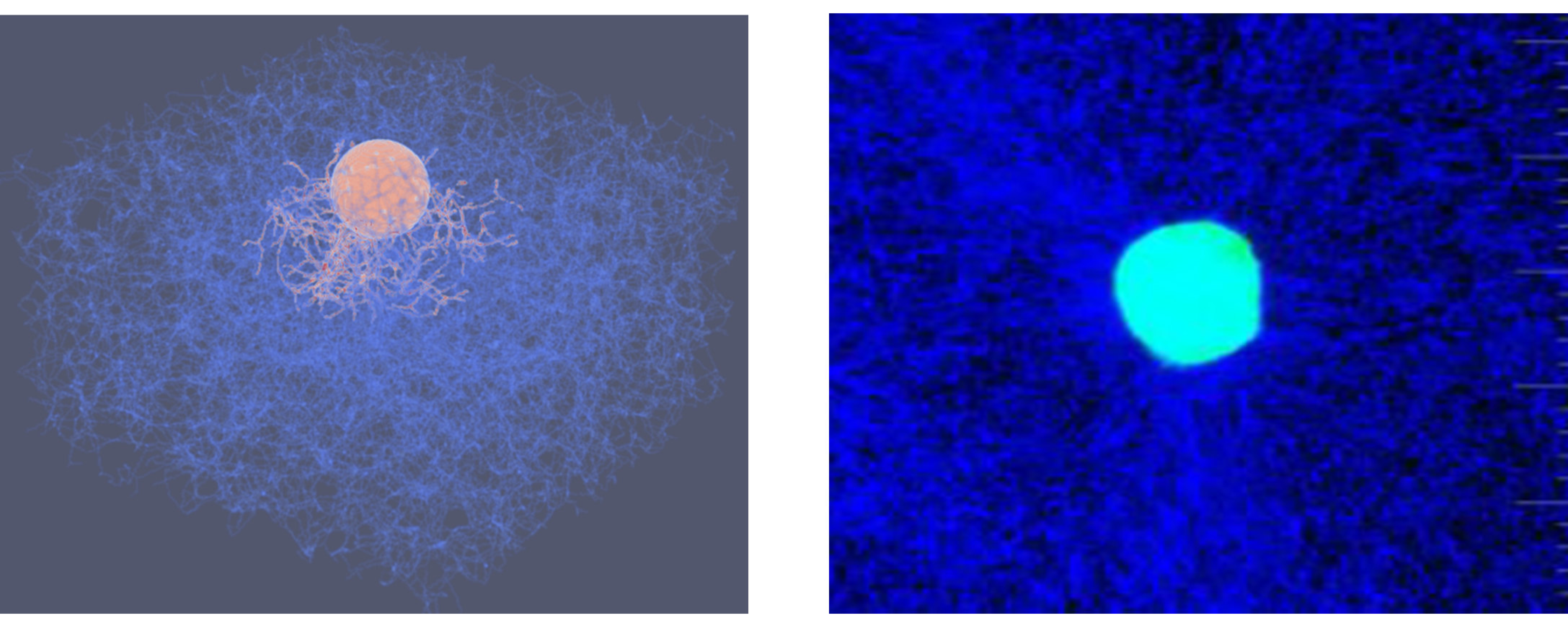}
\caption{Left panel: 3D visualization of a single MCF-10A cell
migrating on isotropic collagen gel with randomly oriented fibers.
The contraction of the actin filaments generates active tensile
forces, which are transmitted to the collagen fibers (shown in red
or dark gray in print version). Right panel: Confocal microscopy
image of a migrating MCF-10A cell (bright blue) on collagen gel.
The collagen fibers are shown in dark blue (or dark gray in print
version). The linear size of the system is $\sim 100 \mu m$.}
\label{fig_simu_iso}
\end{figure}

Figure \ref{fig_simu_iso} shows the 3D visualization of a single
MCF-10A migrating on isotropic collagen gel with randomly oriented
fibers (see the left panel). The 3D collagen network model is
obtained via stochastic reconstruction based on the structural
statistics extracted from confocal images, as described in Sec.
IIA. The contraction of the actin filaments generates active
tensile forces, which are transmitted to the collagen fibers and
propagate in the ECM network. The fibers carrying large tensile
forces are highlighted in red color. The right panel of the figure
shows a confocal microscopy image of a migrating MCF-10A cell
(bright blue) on isotropic collagen gel. It can be seen that the
collagen fibers in the vicinity of the cell surface tend to orient
perpendicularly to the cell surface, implying that the cell
generates traction forces and pulls the fibers, consistent with
the simulation results.

\begin{figure}[htp]
\centering
\includegraphics[width=0.485\textwidth]{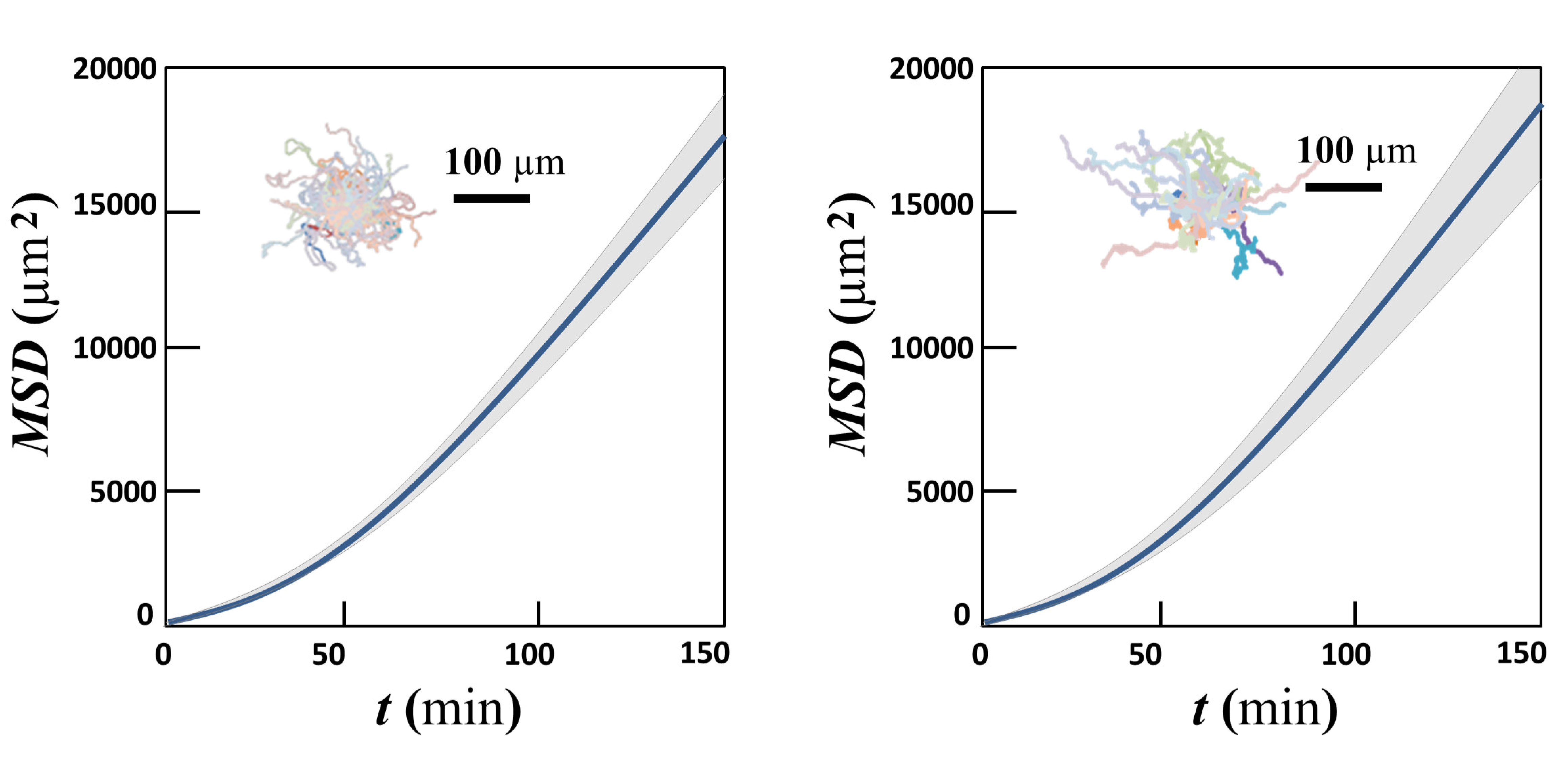}
\caption{Comparison of the mean square displacement (MSD) of a
single MCF-10A cell migrating on isotropic collagen gel with
randomly oriented fibers, respectively obtained from computer
simulation (left panel) and {\it in vitro} experiment (right
panel). The insets show the trajectories of the cells.}
\label{fig_msd_iso}
\end{figure}

Figure \ref{fig_msd_iso} shows the mean square displacement (MSD)
of a single MCF-10A cell migrating on isotropic collagen gel with
randomly oriented fibers, respectively obtained from computer
simulation (left panel) and {\it in vitro} experiment (right
panel). The initial ballistic dynamics (i.e., $\sigma^2 \propto
t^2$) can be clearly seen, which is followed by the diffusive
dynamics (i.e., $\sigma^2 \propto Dt$). The cell diffusivity
obtained from the simulations and experiments are respectively $D
\approx 94 \mu m^2/min$ and $D \approx 103 \mu m^2/min$, which
agree well with one another. The insets of Fig. \ref{fig_msd_iso}
show the trajectories of an ensemble cells respectively obtained
from the simulations and experiments. It can be clearly seen that
the cell migration is isotropic, as expected for a cell in a
homogeneous micro-environment without any externally applied cues.
These results clearly indicate the validity of our model.




\subsection{Migration dynamics of MCF-10A cells on collagen gel with aligned fibers}

\begin{figure}[htp]
\centering
\includegraphics[width=0.385\textwidth]{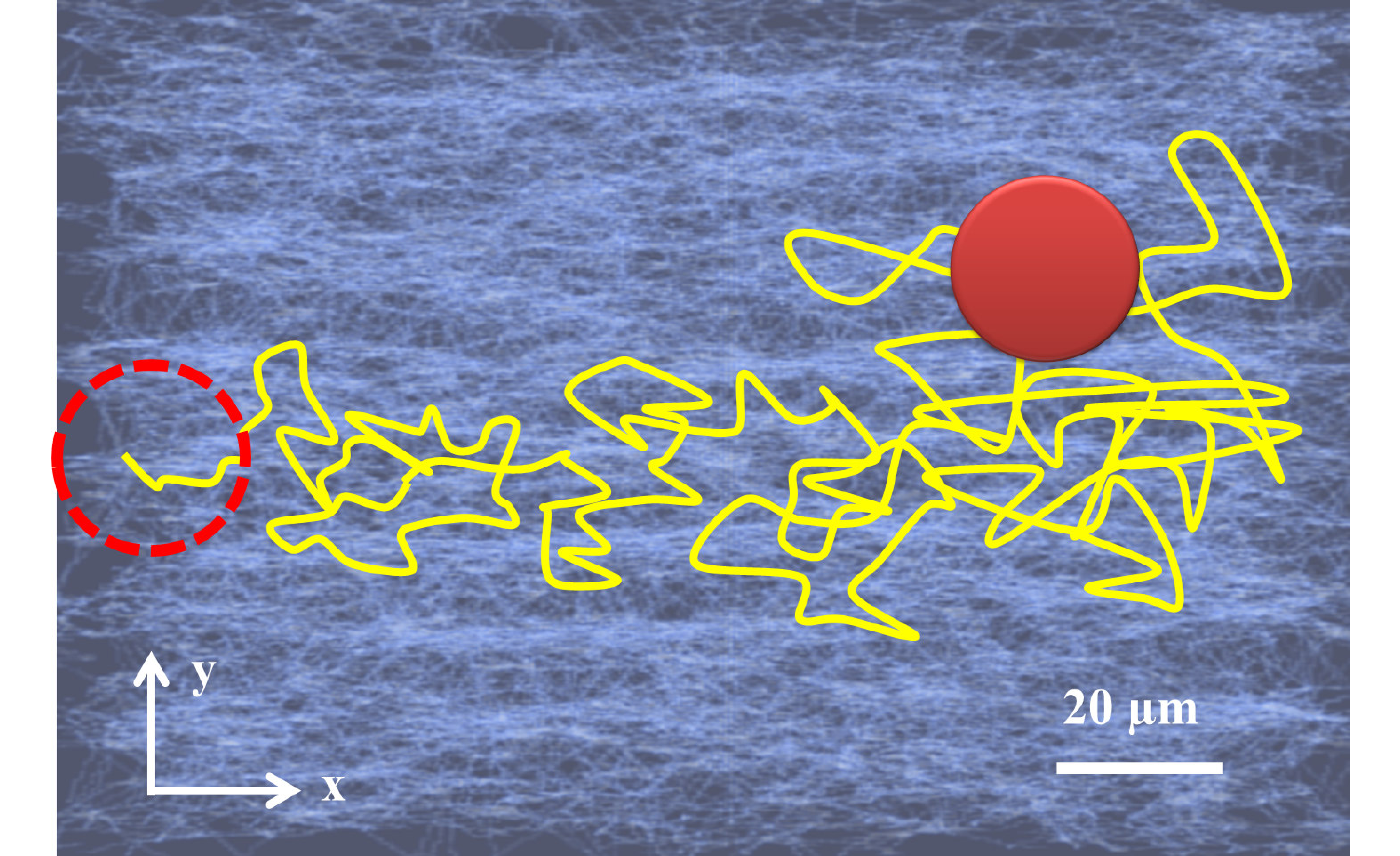}
\caption{A typical trajectory of MCF-10A cell migrating on 3D
collagen gel with horizontally aligned fibers obtained from
simulations.} \label{fig_traj_align}
\end{figure}

With our model validated by experiments, we now employ it to study
cell migration in complex micro-environment, such as collagen gels
with aligned fibers, which are difficult to fabricate
experimentally. The 3D virtual ECM networks are stochastic
constructed by maximizing the fiber orientation metric $\Omega$
along the x-direction (see Sec. IIA for details). This leads to
model networks with fibers mainly aligned along the x-direction
(see Fig. \ref{fig_traj_align}).

\begin{figure}[htp]
\centering
\includegraphics[width=0.485\textwidth]{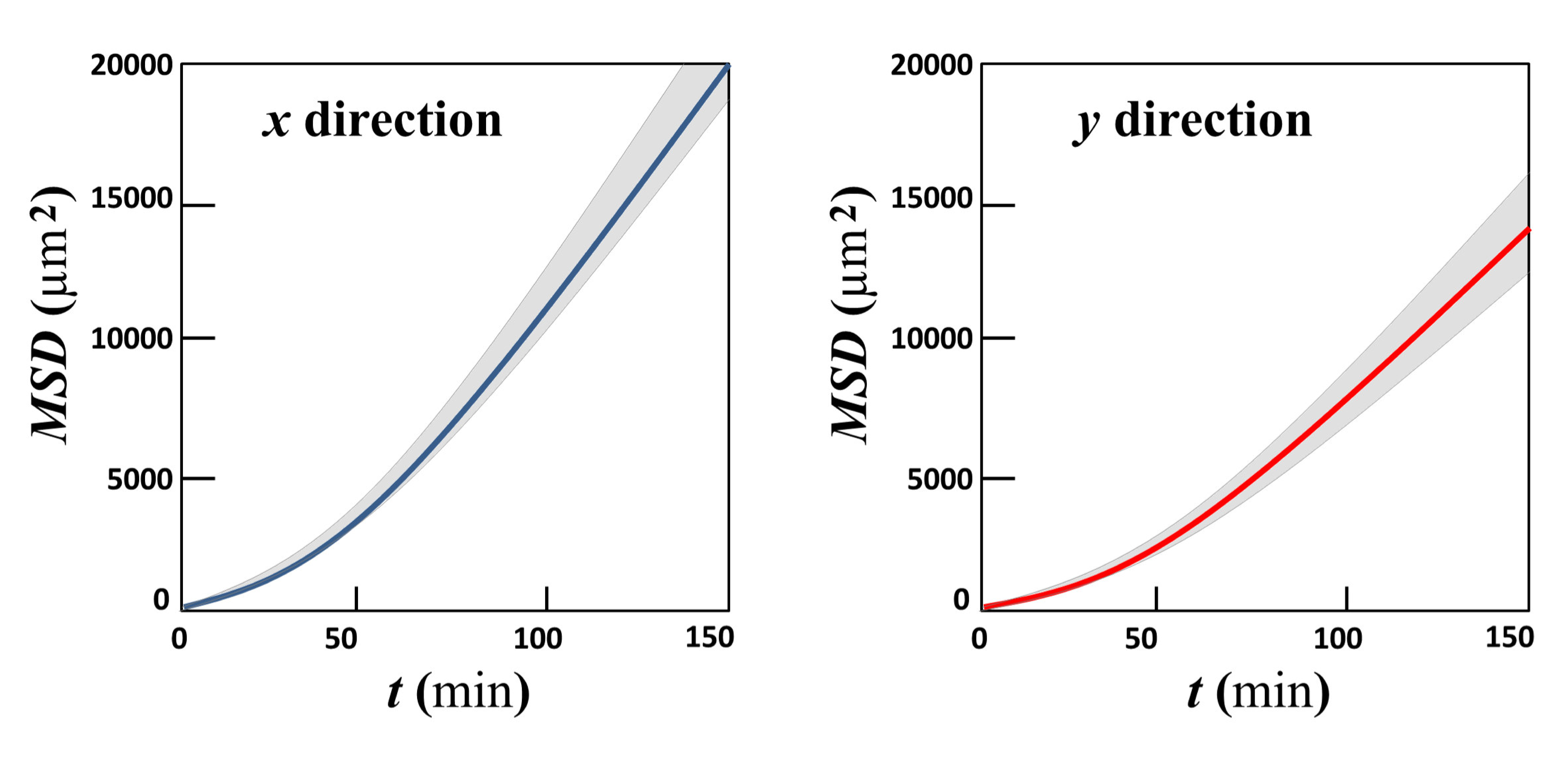}
\caption{Mean squared displacement (MSD) of MCF-10A cell migrating
on 3D collagen gel with horizontally aligned fibers (along the
x-direction) respectively along the x-direction (left panel) and
y-direction (right panel). Anisotropy in migration can be clearly
observed, i.e., the cell tends to move along the direction of
fiber aligned, a phenomenon known as contact guidance.}
\label{fig_msd_align}
\end{figure}

Figure \ref{fig_traj_align} shows a typical trajectory of a
MCF-10A cell migrating on 3D collagen gel with horizontally
aligned fibers obtained from simulations. It can be clearly seen
that the cell tends to migrate along the direction consistent with
the fiber alignment direction (e.g., in this case, x-direction).
This can also been seen quantitatively seen from the MSD analysis.
Figure \ref{fig_msd_align} shows the MSD of the migrating cell
respectively along the x-direction (left panel) and y-direction
(right panel). Anisotropy in the migration can be clearly
observed, i.e., the cell moves much faster long the fiber
alignment direction than the perpendicular direction.

We note that the phenomenon that cells tend to migrate along the
fiber alignment direction is well known and termed as ``contact
guidance'' \cite{ref18, ref19}. In our simulations, as the
migrating cell pulls the ECM fibers, the large tensile forces are
mainly carried by chains of aligned fibers, which are typically
referred to as ``force chains'' \cite{ref11, ref33}. The
high-stress fibers on the force chains are effective stiffer
(e.g., due to strain hardening) and thus, can support
large-magnitude locomotion steps along chain directions, and in
this case, the fiber alignment direction.




\subsection{Migration dynamics of MCF-10A cells on collagen gel with a stiffness gradient}

We now employ our model to study cell migration dynamics on
collagen gels with a stiffness gradient. As described in Sec. IIA,
the structural model of the 3D ECM is constructed based on the
experimentally obtained statistics of a 2 mg/ml collagen gel with
randomly oriented fibers. Once the 3D structural model is
obtained, a linear stiffness distribution a long x-direction with
a constant gradient is built. This is achieved by re-scaling the
Young's modulus of the fiber according to $E(x) = E\cdot (1 +
x/L)$, where $x$ is the x-coordinate of the center of the fiber.

\begin{figure}[htp]
\centering
\includegraphics[width=0.385\textwidth]{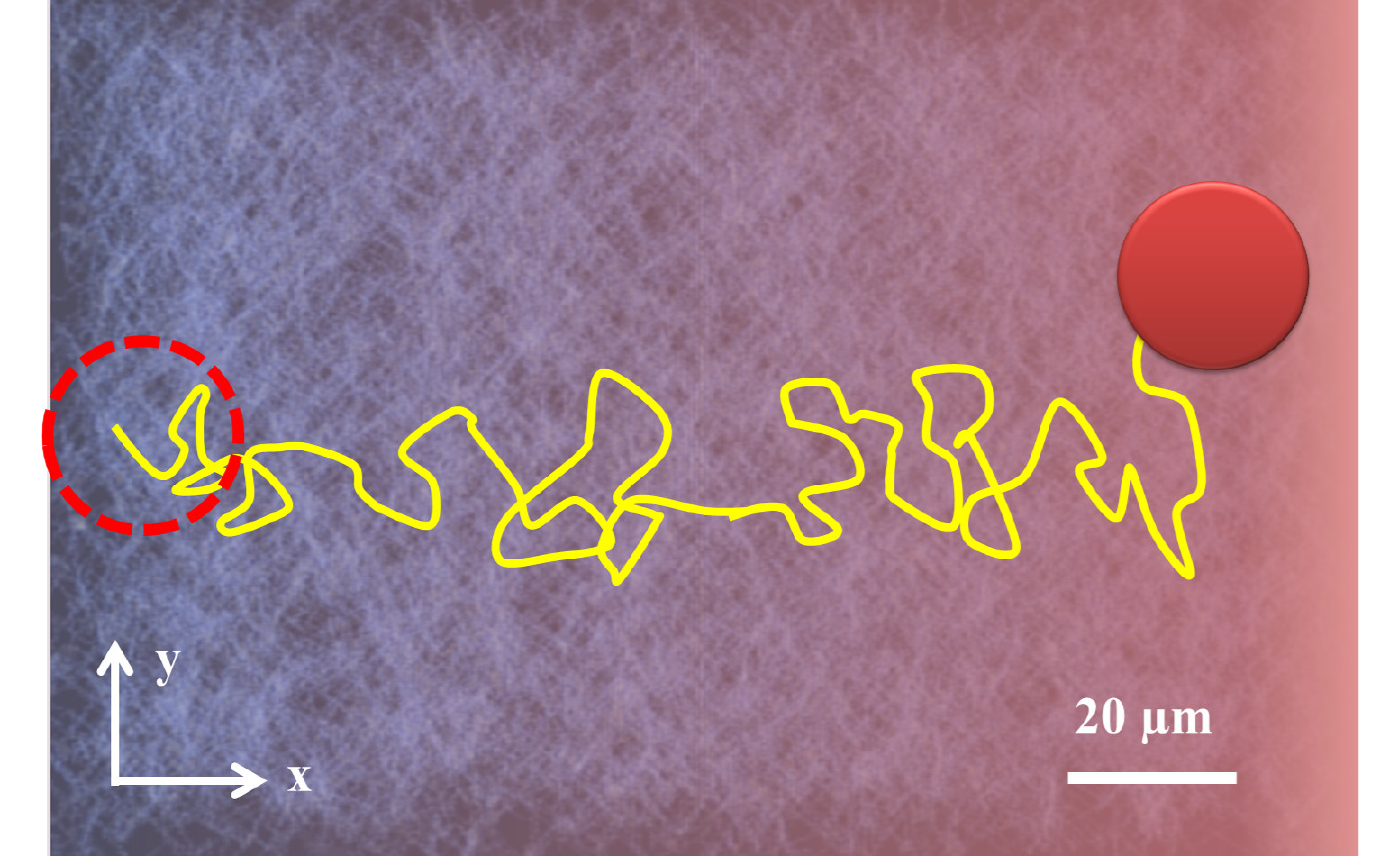}
\caption{A typical trajectory of MCF-10A cell migrating on 3D
collagen gel with a stiffness gradient along the x-direction
obtained from simulations.} \label{fig_traj_duro}
\end{figure}

\begin{figure}[htp]
\centering
\includegraphics[width=0.485\textwidth]{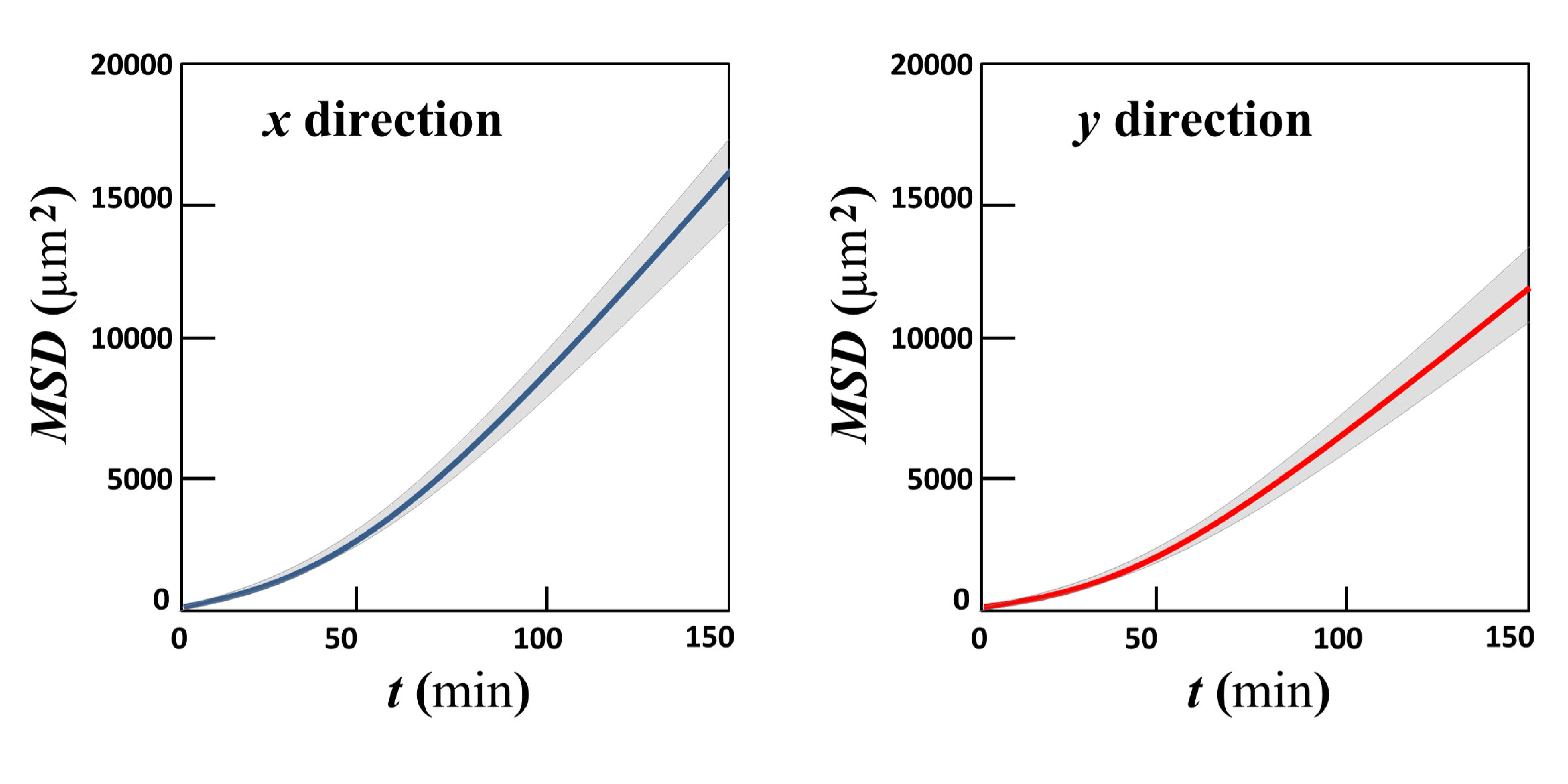}
\caption{Mean squared displacement (MSD) of MCF-10A cell migrating
on 3D collagen gel with a stiffness gradient (along the
x-direction), respectively along the x-direction (left panel) and
y-direction (right panel). Anisotropy in migration can be clearly
observed, i.e., the cell tends to move up against the stiffness
gradient, a phenomenon known as durotaxis.} \label{fig_msd_duro}
\end{figure}

Figure \ref{fig_traj_duro} shows a typical trajectory of a MCF-10A
cell migrating on 3D collagen gel with a stiffness gradient along
the x-direction. Similar to the case of contact guidance, it can
be clearly seen that the cell tends to migrate along the direction
against the stiffness gradient along the positive x-direction.
This can also been seen quantitatively seen from the MSD analysis.
Figure \ref{fig_msd_duro} shows the MSD of the migrating cell
respectively along the x-direction (left panel) and y-direction
(right panel). Anisotropy in the migration can be clearly
observed, i.e., the cell moves much faster long the stiffness
gradient direction than the perpendicular direction. We note that
an important distinction between migration anisotropy in this case
and the contact guidance case is that here the cell migration is
uni-directional, i.e., up the stiffness gradient; while in the
contact guidance case, the migration is bi-directional, i.e.,
along the fiber alignment direction but the cells can go in both
ways.

The phenomenon that cells migrate against stiffness gradient of
the ECM is well known and termed as ``durotaxis'' \cite{ref15,
ref16, Brown2009}. In our simulations, as the migrating cell pulls
the ECM fibers, the stiffer fibers will possess smaller
deformation, which in turn leads to larger locomotion components
towards these fibers (c.f. Eq.(\ref{eq_fila})). The accumulated
effect of many local migration steps is the overall biased
migration up the stiffness gradient as observed in the
experiments.




\section{Strongly correlated multi-cellular dynamics}

In Sec. IV, we show that our computational model can capture the
salient features of single-cell migration dynamics in both
homogeneous and complex micro-environment. In this section, we
employ the model to investigate multi-cellular migration dynamics.
As mentioned in Sec. III, we do not explicitly model cell-cell
adhesion here (due to the weak adhesion between the cancer cells)
and use a minimal model for cell-cell repulsion due to contact
inhibition (see Sec. III for details). In addition, in this study,
we focus on relatively small system, containing $\sim 20$ cells.

\begin{figure}[htp]
\includegraphics[width=0.485\textwidth]{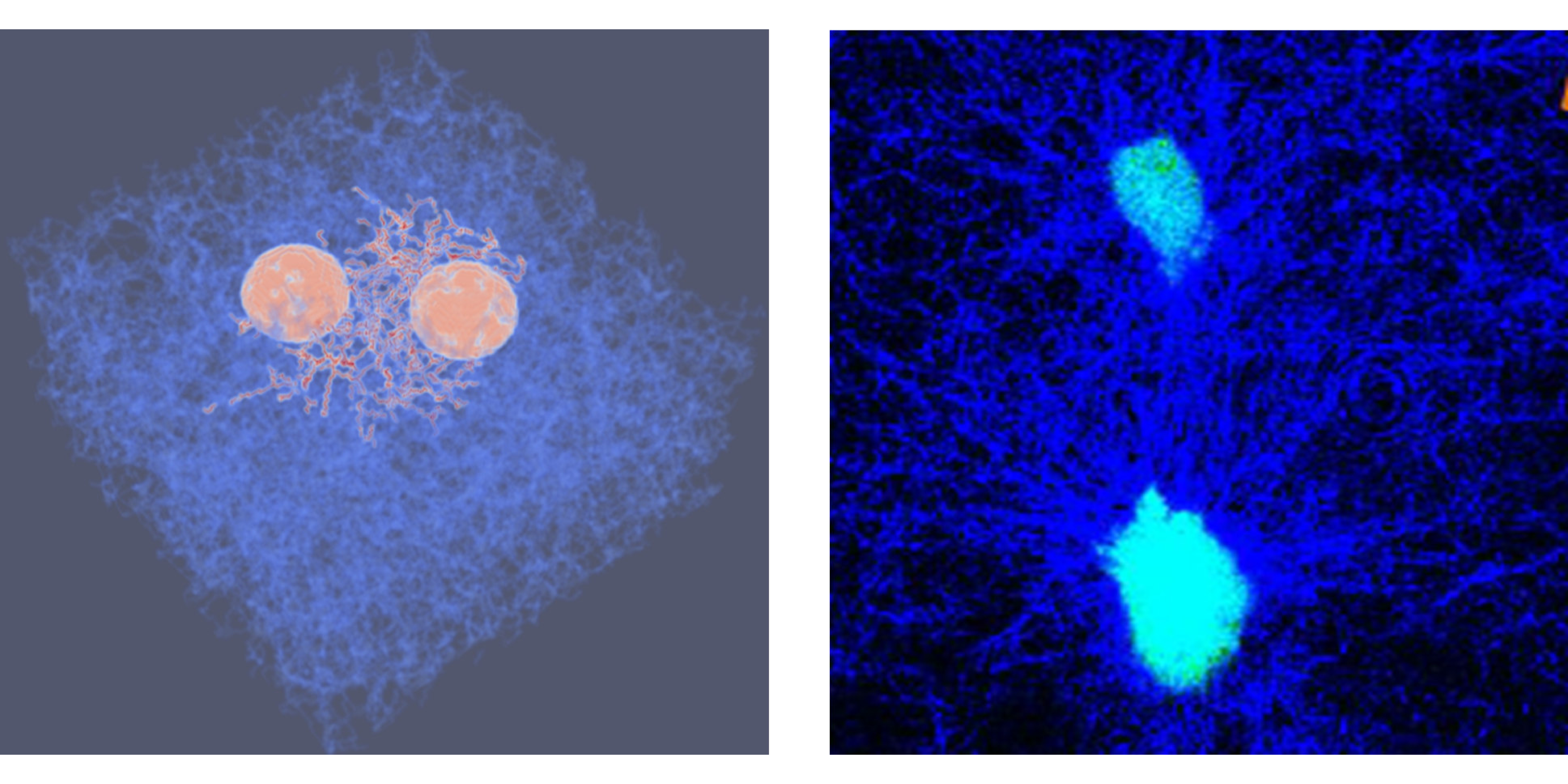}
\caption{Left panel: 3D visualization of two closely spaced
MCF-10A cells migrating on isotropic collagen gel with randomly
oriented fibers. The collagen fibers carrying large tensile forces
generated by actin filament contraction are highlighted in red.
Right panel: Confocal microscopy image of a pair of migrating
MCF-10A cells (bright blue) on collagen gel. The collagen fibers
are shown in dark blue (or dark gray in print version).}
\label{fig_simu_mult}
\end{figure}

Figure \ref{fig_simu_mult}(a) shows 3D visualization of a small
portion (with a linear size $\sim 50 \mu m$) of the simulation box
which contains two closely spaced MCF-10A cells migrating on
isotropic collagen gel with randomly oriented fibers. The active
tensile forces generated by the cells (due to actin filament
contraction) are transmitted to the collagen fibers. The collagen
fibers carrying large tensile forces are highlighted in red.
Figure \ref{fig_simu_mult}(a) shows the confocal microscopy image
of a pair of migrating MCF-10A cells (bright blue) on collagen
gel. It can be clearly seen that the collagen fibers (dark blue)
between the two cells form a mesoscopic scale structure, which is
clearly distinguished from original homogeneous ECM network and is
consistent with the meso-scale structure formed by the high-stress
fibers in our simulations.

\begin{figure}[htp]
\centering
\includegraphics[width=0.4\textwidth]{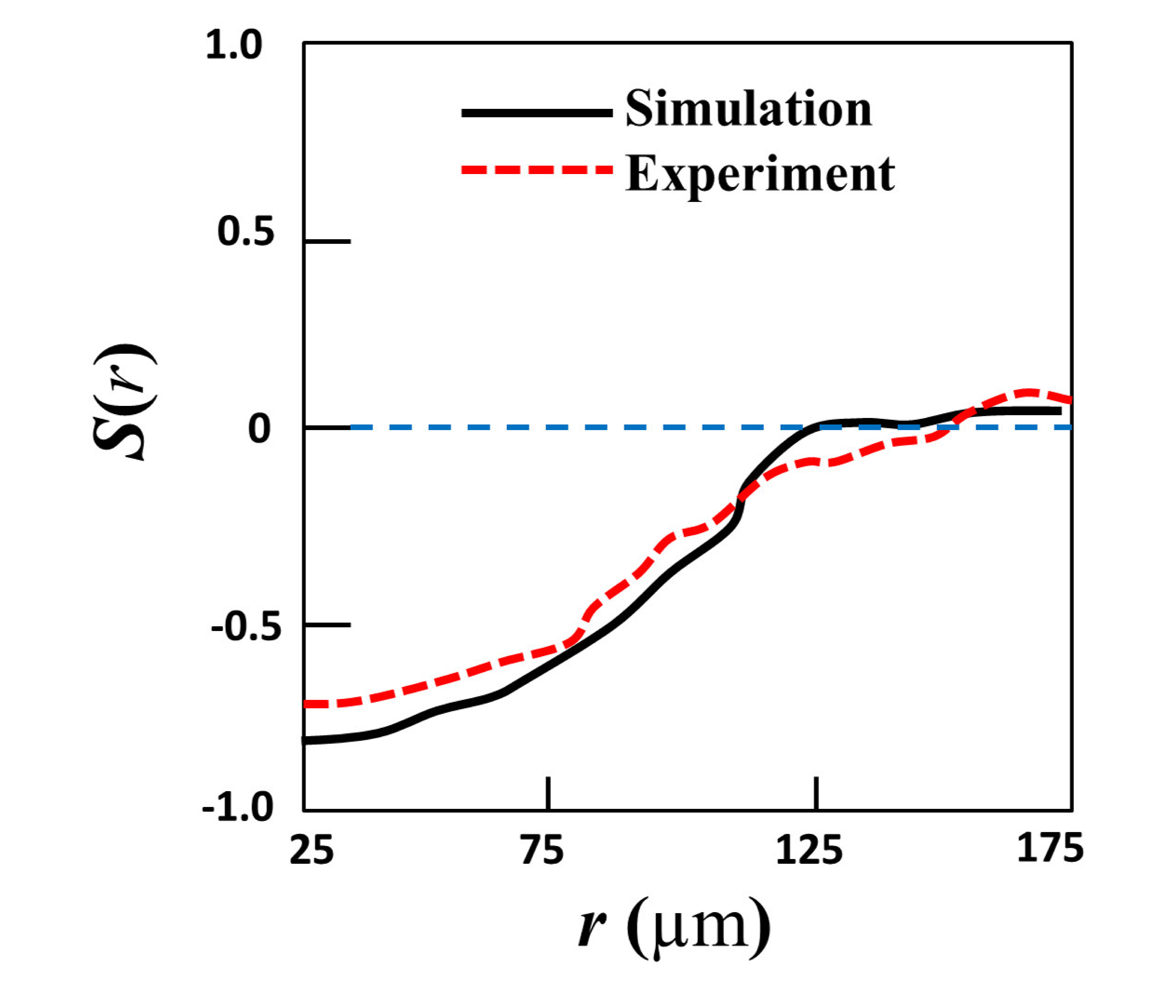}
\caption{Comparison of the velocity correlation function $S(r)$
(see the text for definition) of MCF-10A cells migrating on
isotropic collagen gel with randomly oriented fibers, respectively
obtained from computer simulation (solid curve) and {\it in vitro}
experiment (dashed curve).} \label{fig_corr_mult}
\end{figure}


To quantify the correlations in the collective migration dynamics
of multiple MCF-10A cells, we employ the velocity correlation
function $S(r)$, i.e.,
\begin{equation}
S(r) = <{\bf v}_i({\bf x})\cdot{\bf v}_j({\bf x}+{\bf r})>/(|{\bf
v}_i({\bf x})||{\bf v}_j({\bf x}+{\bf r})|)
\end{equation}
where $r = |{\bf r}|$, $i$, $j$ denote a pair of cells connected
by the remodeled meso-scale ECM structures and $<, >$ denotes
ensemble average over all different cell pairs. We note that in
computing $S(r)$, we only consider a subsets of cell pairs, i.e.,
those between which the meso-scale structures are formed. This
allows us to clearly obtain the effects of such meso-structure on
the collective dynamics of the cells, if any. Due to cell's mutual
exclusion effects, $S(r) = 0$ for $r<D^*$ and $D^*$ is roughly the
diameter of a cell. In addition, two cells separated by very large
distances are not correlated, i.e., $S(r) \approx 0$ for large $r$
values. A positive $S(r)$ indicates that the cells tend to move in
the same direction, implying a net ``flow'' of cells in the
system. On the other hand, a negative $S(r)$ indicates that the
cells move towards one another, implying the formation of
aggregation or clusters.

Figure \ref{fig_corr_mult} shows the velocity correlation function
$S(r)$ of MCF-10A cells migrating on isotropic collagen gel with
randomly oriented fibers, respectively obtained from computer
simulation (solid curve) and {\it in vitro} experiment (dashed
curve). It can be seen that the simulation results agree very well
with experimental data. Interestingly, the $S(r)$ functions
(beyond the trivial exclusion region) start from a very negative
value (close to the minimal value -1) around $D^* \approx 25 \mu
m$, slowly increase to zero and then fluctuate around zero. This
indicates the cells tend to move towards one another, facilitated
by the mesoscopic structures of the remodeled ECM, which is also
confirmed by time-lapse confocal data \cite{qihui2019}.


Our results indicate that strongly correlated cell migration
dynamics is correlated with the meso-scale ECM structures due to
cell remodeling. One possible reason is that the meso-structures
are composed of many force chains (or a ``force network''), which
are in turn composed of fibers carrying large tensile forces.
Therefore, the fibers in the meso-structures (at least in our
simulations) are stiffer than the remaining stress-free fibers,
which indicates that the meso-structures themselves are stiffer
than the surrounding ECM. These stiffer meso-structures can then
facilitate correlated cell migration via durotaxis, and also
facilitate indirect mechanical coupling between the migrating
cells.






\section{Conclusions and discussion}

In this paper, we develop a computational model for cell migration
in complex micro-environment, which explicitly takes into account
a variety of cellular level processes including focal adhesion
formation and disassembly, active traction force generation and
cell locomotion due to actin filament contraction, transmission
and propagation of tensile forces in the ECM. We employ
statistical descriptors obtainable from confocal imagining to
quantify and control the 3D ECM network microstructure and use a
nonlinear mechanical model for the ECM networks, which
incorporates buckling of collagen fibers upon compression and
strain-hardening upon stretching. We validate our model by
accurately reproducing single-cell dynamics of MCF-10A breast
cancer cells migrating on collagen gels and show that the
durotaxis and contact guidance effects naturally arise as a
consequence of the cell-ECM micro-mechanical interactions
considered in the model. Moreover, our model predicts strongly
correlated multi-cellular migration dynamics, which are resulted
from the ECM-mediated mechanical coupling among the migrating cell
and are subsequently verified in {\it in vitro} experiments using
MCF-10A cells.


Although focusing on the non-metastatic MCF-10A breast cancer
cells migrating on 3D collagen gels, our model can be generalize
to investigate the migration of mesenchymal cells (e.g., invasive
MDA-MB-231 breast cancer cells) in 3D ECM. The key modification is
to explicitly model ECM degradation by the cells, which can be
achieved using the following rule: A migrating cell degrades
collagen fibers with a probability $p_b\propto \exp(-r)$, with $r$
the distance from the fiber to the cell membrane. A degraded fiber
is removed from the network in subsequent simulation steps. In
addition, cell-cell adhesion can also be easily incorporated into
the model to investigate a wide range of cell lines with different
phenotypes. With proper modifications and generalizations, as well
as efficient parallel implementation, it is expected that the
model could be employed to investigate collective migratory
behaviors and emergent self-organizing multi-cellular patterns
resulted from ECM-mediated mechanical signaling among the cells.



\begin{acknowledgments}
Y. Z., H. N. and Y. J. thank Arizona State University for the
generous start-up funds and the University Graduate Fellowships.
Q. F., X. W., and F. Y. thank the Chinese Academy of Sciences
(CAS), the Key Research Program of Frontier Sciences of CAS (Grant
No. QYZDB-SSW-SYS003). L. L. and R. L. thank the National Natural
Science Foundation of China (Grants No. 11474345, No. 11674043,
No. 11604030, and No. 11774394). B. S. thank the support from the
Scialog Program sponsored jointly by Research Corporation for
Science Advancement and the Gordon and Betty Moore Foundation. B.
S. is partially supported by the Medical Research Foundation of
Oregon and SciRIS-II award from Oregon State University and by the
National Science Foundation Grant PHY-1400968.

\end{acknowledgments}


\begin{thebibliography}{44}



\bibitem{ref1}
A. Aman, and T. Piotrowski, Cell migration during morphogenesis.
Developmental Biology {\bf 341}, 20-33 (2010)

\bibitem{ref2}
P. Friedl, and D. Gilmour, Collective cell migration in
morphogenesis, regeneration and cancer. Nature Reviews Molecular
Cell Biology {\bf 10}, 445 (2009).

\bibitem{ref3}
A. Vaezi, C. Bauer, V. Vasioukhin, and E. Fuchs, Actin cable
dynamics and Rho/Rock orchestrate a polarized cytoskeletal
architecture in the early steps of assembling a stratified
epithelium. Developmental Cell {\bf 3}, 367-381 (2002).

\bibitem{ref4}
S. Werner, T. Krieg, and H. Smola, Keratinocyte–fibroblast
interactions in wound healing. Journal of Investigative
Dermatology {\bf 127}, 998-1008 (2007).

\bibitem{ref12}
A. J. Ridley, M.A. Schwartz, K. Burridge, R.A. Firtel, M.H.
Ginsberg, G. Borisy, J.T. Parsons, and A.R. Horwitz, Cell
migration: integrating signals from front to back. Science {\bf
302}, 1704-1709 (2003).

\bibitem{ref13}
P. Friedl, and E.-B. Brocker, The biology of cell locomotion
within three-dimensional extracellular matrix. Cellular and
Molecular Life Sciences CMLS {\bf 57}, 41-64 (2000).

\bibitem{ref14}
H. Szurmant, and G.W. Ordal, Diversity in chemotaxis mechanisms
among the bacteria and archaea. Microbiology and Molecular Biology
Reviews {\bf 68}, 301-319 (2004).

\bibitem{ref15}
S. V. Plotnikov, A.M. Pasapera, B. Sabass, and C.M. Waterman,
Force fluctuations within focal adhesions mediate ECM-rigidity
sensing to guide directed cell migration. Cell {\bf 151},
1513-1527 (2012).

\bibitem{ref16}
R. Sunyer, V. Conte, J. Escribano, A. Elosegui-Artola, A.
Labernadie, L. Valon, D. Navajas, J.M. García-Aznar, J.J. Munoz,
and P. Roca-Cusachs, Collective cell durotaxis emerges from
long-range intercellular force transmission. Science {\bf 353},
1157-1161 (2016).

\bibitem{Brown2009}
E. Hadjipanayi, V. Mudera, and R. A. Brown, Guiding cell migration
in 3D: a collagen matrix with graded directional stiffness. {Cell
Motil. Cytoskeleton} {\bf 66} 121-8 (2009).

\bibitem{ref17}
S. B. Carter, Haptotaxis and the mechanism of cell motility.
Nature {\bf 213}, 256 (1967).

\bibitem{ref18}
P. P. Provenzano, D.R. Inman, K.W. Eliceiri, S.M. Trier, and P.J.
Keely, Contact guidance mediated three-dimensional cell migration
is regulated by Rho/ROCK-dependent matrix reorganization.
Biophysical Journal {\bf 95}, 5374-5384 (2008).

\bibitem{ref19}
J. H. Wang, and E.S. Grood, The strain magnitude and contact
guidance determine orientation response of fibroblasts to cyclic
substrate strains. Connective Tissue Research {\bf 41}, 29-36
(2000).

\bibitem{Tranquillo1993}
S. Guido and R. T. Tranquillo, A methodology for the systematic
and quantitative study of cell contact guidance in oriented
collagen gels. Correlation of fibroblast orientation and gel
birefringence. {J. Cell Sci.} {\bf 105}, 317-31 (1993).


\bibitem{ref20}
S. Wang, and P.G. Wolynes, Active contractility in actomyosin
networks. Proceedings of the National Academy of Sciences {\bf
109}, 6446-6451 (2012).

\bibitem{ref21}
T. Lecuit, P.-F. Lenne, and E. Munro, Force generation,
transmission, and integration during cell and tissue
morphogenesis. Annual Review of Cell and Developmental Biology
{\bf 27}, 157-184 (2011).

\bibitem{ref22}
M. A. Schwartz, Integrins and extracellular matrix in
mechanotransduction. Cold Spring Harbor perspectives in biology,
a005066 (2010).

\bibitem{ref23}
G. Totsukawa, Y. Wu, Y. Sasaki, D.J. Hartshorne, Y. Yamakita, S.
Yamashiro, and F. Matsumura, Distinct roles of MLCK and ROCK in
the regulation of membrane protrusions and focal adhesion dynamics
during cell migration of fibroblasts. The Journal of Cell Biology
{\bf 164}, 427-439 (2004).

\bibitem{ref24}
C. A. Jones, M. Cibula, J. Feng, E.A. Krnacik, D.H. McIntyre, H.
Levine, and B. Sun, Micromechanics of cellularized biopolymer
networks. Proceedings of the National Academy of Sciences {\bf
112}, E5117-E5122 (2015).

\bibitem{ref25}
S. B. Lindstrom, D.A. Vader, A. Kulachenko, and D.A. Weitz,
Biopolymer network geometries: characterization, regeneration, and
elastic properties. Physical Review E {\bf 82}, 051905 (2010).

\bibitem{ref26}
H. Mohammadi, P.D. Arora, C.A. Simmons, P.A. Janmey, and C.A.
McCulloch, Inelastic behaviour of collagen networks in cell–matrix
interactions and mechanosensation. Journal of The Royal Society
Interface {\bf 12}, 20141074 (2015).

\bibitem{ref27}
S. Nam, K.H. Hu, M.J. Butte, and O. Chaudhuri, Strain-enhanced
stress relaxation impacts nonlinear elasticity in collagen gels.
Proceedings of the National Academy of Sciences, 201523906 (2016).

\bibitem{ref28}
Nam, S., J. Lee, D.G. Brownfield, and O. Chaudhuri,
Viscoplasticity enables mechanical remodeling of matrix by cells.
Biophysical journal, 2016. 111(10): p. 2296-2308.

\bibitem{ref29}
J. Kim, J. Feng, C.A. Jones, X. Mao, L.M. Sander, H. Levine, and
B. Sun, Stress-induced plasticity of dynamic collagen networks.
Nature Communications {\bf 8}, 842 (2017).

\bibitem{shaohua2019}
S. Chen, W. Xu, J. Kim, H. Nan, Y. Zheng, B. Sun, and Y. Jiao,
Novel inverse finite-element formulation for reconstruction of
relative local stiffness in heterogeneous extra-cellular matrix
and traction forces on active cells. Physical Biology {\bf 16},
036002 (2019).

\bibitem{ref30}
A. D. Doyle, N. Carvajal, A. Jin, K. Matsumoto, and K.M. Yamada,
Local 3D matrix microenvironment regulates cell migration through
spatiotemporal dynamics of contractility-dependent adhesions.
Nature Communications {\bf 6}, 8720 (2015).

\bibitem{Frey07}
C. Heussinger and E. Frey, Force distributions and force chains in
random stiff fiber networks. {Eur. Phys. J. E} {\bf 24} 47–53
(2007).

\bibitem{ref5}
F. Grinnell, and W.M. Petroll, Cell motility and mechanics in
three-dimensional collagen matrices. Annual Review of Cell and
Developmental Biology {\bf 26}, 335-361 (2010).

\bibitem{ref6}
Y. L. Han, P. Ronceray, G. Xu, A. Malandrino, R.D. Kamm, M. Lenz,
C.P. Broedersz, and M. Guo, Cell contraction induces long-ranged
stress stiffening in the extracellular matrix. Proceedings of the
National Academy of Sciences {\bf 115}, 4075-4080 (2018).

\bibitem{ref7}
X. Ma, M.E. Schickel, M.D. Stevenson, A.L. Sarang-Sieminski, K.J.
Gooch, S.N. Ghadiali, and R.T. Hart, Fibers in the extracellular
matrix enable long-range stress transmission between cells.
Biophysical Journal {\bf 104}, 1410-1418 (2013).

\bibitem{ref8}
P. Ronceray, C.P. Broedersz, and M. Lenz, Fiber networks amplify
active stress. Proceedings of the National Academy of Sciences
{\bf 113}, 2827-2832 (2016).

\bibitem{ref9}
H. Wang, A. Abhilash, C.S. Chen, R.G. Wells, and V.B. Shenoy,
Long-range force transmission in fibrous matrices enabled by
tension-driven alignment of fibers. Biophysical Journal {\bf 107},
2592-2603 (2014).

\bibitem{ref10}
F. Beroz, L.M. Jawerth, S. Munster, D.A. Weitz, C.P. Broedersz,
and N.S. Wingreen, Physical limits to biomechanical sensing in
disordered fibre networks. Nature Communications {\bf 8}, 16096
(2017).

\bibitem{ref11}
L. Liang, C. Jones, S. Chen, B. Sun, and Y. Jiao, Heterogeneous
force network in 3D cellularized collagen networks. Physical
Biology {\bf 13}, 066001 (2016).



\bibitem{model01}
M. H. Zaman, R. D. Kamm, P. Matsudaria, and D. A. Lauffenburger.
Computational model for cell migration in three-dimensional
matrices. Biophys. J. {\bf 89}, 1389 (2005).

\bibitem{model02}
A. Vaziri and A. Gopinath. Cell and biomolecular mechanics in
silico. Nature Materials {\bf 7}, 15 (2008).

\bibitem{model03}
P. Masuzzo, M. Van Troys, C. Ampe, and L. Martens, Taking aim at
moving targets in computational cell migration. Trends in cell
biology {\bf 26}, 88 (2016).

\bibitem{model04}
F. Ziebert, S. Swaminathan, and I. S. Aranson. Modeling for
self-polarization and motility of keratocyte fragments. J. R. Soc.
Interface {\bf 9}, 1084 (2011).

\bibitem{model05}
D. Shao, W. J. Rappel, and H. Levine. Computational model for cell
morphodynamics. Phys. Rev. Lett. {\bf 105}, 108104 (2010).

\bibitem{model06}
D. Shao, H. Levine, and W. J. Rappel. Coupling actin flow,
adhesion, and morphology in a computational cell motility model.
Proc. Natl. Acad. Sci. USA {\bf 109}, 6851 (2012).

\bibitem{model07}
U. Z. George, A. Stephanou, and A. Madzvamuse. Mathematical
modeling and numerical simulations of actin dynamics in the
eukaryotic cell. J. Math. Biol. {\bf 66}, 547 (2013).

\bibitem{model08}
T. C. Bidone, W. Jung, D. Maruri, C. Borau, R. D. Kamm, and T.
Kim, Morphological transformation and force generation of active
cytoskeletal networks, PLoS Comput. Biol. {\bf 13}, e1005277
(2017).

\bibitem{model09}
M. C. Kim, J. Whisler, Y. R., Silberberg, et al. Cell invasion
dynamics into a three dimensional extracellular matrix fibre
network. PLoS Comput. Biol. {\bf 11}, e1004535 (2015).



\bibitem{model10}
T. Vicsek, A. Czirok, E. Ben-Jacob, I. Cohen, and O. Shochet,
Novel type of phase transition in a system of self-driven
particles. Physical review letters {\bf 75}, 1226 (1995).

\bibitem{model11}
C. Bechinger, R. Di Leonardo, H. Lowen, C. Reichhardt, G. Volpe,
Active particles in complex and crowded environments. Reviews of
Modern Physics {\bf 88}, 045006 (2016).

\bibitem{model12}
D. Bi, J. Lopez, J. Schwarz, M. L. Manning, A density-independent
rigidity transition in biological tissues, Nature Physics {\bf
11}, 1074 (2015).

\bibitem{model13}
F. Graner, and J. A. Glazier, Simulation of biological cell
sorting using a two-dimensional extended Potts model. Physical
Review Letters {\bf 69}, 2013 (1992).

\bibitem{model14}
Y. Jiao and S. Torquato, Emergent Properties from a Cellular
Automaton Model for Invasive Tumor Growth in Heterogeneous
Environment, PLoS Computational Biology {\bf 7}, 1002314 (2011).

\bibitem{model15}
Y. Jiao and S. Torquato, Diversity of Dynamics and Morphologies of
Invasive Solid Tumors, AIP Advances {\bf 2}, 011003 (2012)

\bibitem{model16}
Y. Jiao and S. Torquato, Evolution and Morphology of
Microenvironment-Enhanced Malignancy of Three-Dimensional Invasive
Solid Tumors, Physical Review E {\bf 87}, 052707 (2013)

\bibitem{model17}
H. Xie, Y. Jiao, Q. Fan, et. al., Modeling Three-dimensional
Invasive Solid Tumor Growth in Heterogeneous Microenvironment
under Chemotherapy, PLoS One {\bf 13}, e0206292 (2018)

\bibitem{model18}
H. Abdel-Rahman, B. Thomas, and T. Kim, A mechanical model for
durotactic cell migration, ACS Biomater Sci Eng (2019)

\bibitem{ref31}
M. Dietrich, H. Le Roy, D. B. Bruckner, H. Engelke, R. Zantl, J.
O. Radler and C. P. Broedersz, Guiding 3D cell migration in
deformed synthetic hydrogel microstructures, Soft Matter {\bf 14},
2816 (2018).


\bibitem{ref33}
Nan, H., L. Liang, G. Chen, L. Liu, R. Liu, and Y. Jiao,
Realizations of highly heterogeneous collagen networks via
stochastic reconstruction for micromechanical analysis of tumor
cell invasion. Physical Review E {\bf 97}, 033311 (2018).

\bibitem{sim_annealing}
S. Kirkpatrick, C. D. Gelatt, and M. P. Vecchi, Optimization by
simulated annealing. Science {\bf 220}, 671 (1983).



\bibitem{MacKintosh05}
D. A. Head, A. J. Levine, and F. C. MacKintosh, Mechanical
response of semiflexible networks to localized perturbations.
{Phys. Rev. E} {\bf 72} 061914 (2005).

\bibitem{Safran12}
Y. Shokef and S. A. Safran, Scaling laws for the response of
nonlinear elastic media with implications for cell mechanics. {\it
Phys. Rev. Lett.} {\bf 108}, 178103 (2012).

\bibitem{nat_method15}
J. Steinwachs, C. Metzner, K. Skodzek, et. al. Three-dimensional
force microscopy of cells in biopolymer networks. {\it Nat.
Method} {\bf 13} 171-6 (2016).

\bibitem{Zallen2009}
R. Fernandez-Gonzalez, M. S. Simoes, J. C. Roper, S. Eaton, and J.
A. Zallen. Myosin ii dynamics are regulated by tension in
intercalating cells. Dev. Cell {\bf 17}, 736–743 (2009).

\bibitem{Munro2011}
T. Lecuit, P. Lenne, and E. Munro. Force generation, transmission
and integration during cell and tissue morphogenesis. Annu. Rev.
Cell Dev. Biol. {\bf 27}, 157–184 (2011).

\bibitem{Matsumura2004}
G. Totsukawa, Y. Wu, Y. Sasaki, D. J. Hartshorne, Y. Yamakita, S.
Yamashiro, and F. Matsumura. Distinct roles of mlck and rock in
the regulation of membrane protrusions and focal adhesion dynamics
during cell migration of fibroblasts. J. Cell Biol. {\bf 164},
427–439 (2004).

\bibitem{boey1998}
S. K. Boey, D. H. Boal, and D. Discher. Simulations of the
erythrocyte cytoskeleton at large deformation. i. microscopic
models. Biophys. J. {\bf 75}, 1573–1583 (1998).

\bibitem{boey1998II}
D. Discher, D. H. Boal, and S. K. Boey. Simulations of the
erythrocyte cytoskeleton at large deformation. ii. micropipette
aspiration. Biophys. J. {\bf 75}, 1584–1597 (1998).

\bibitem{coughlin03}
M. F. Coughlin and D. Stamenovic. A prestressed cable network
model for the adherent cell cytoskeleton. Biophys. J. {\bf 84},
1328–1336 (2003).

\bibitem{gordon2012}
D. Gordon, A. Bernheim-Groswasser, C. Keasar, and O. Farago.
Hierarchical self-organization of cytoskeletal active networks.
Phys. Biol. {\bf 9}, 026005 (2012).

\bibitem{qihui2019}
Q. Fan, Y. Zheng, H. Nan, Y. Jiao, and F. Ye, Strongly correlated
cell dynamics induced by ECM-mediated long-range mechanical
coupling, submitted.

\bibitem{RMP}
C. Bechinger, R. D. Leonardo, H. Lowen, C. Reichhardt, G. Volpe
and G. Volpe, Active particles in complex and crowded
microenvironments, Rev. Mod. Phys. {\bf 88}, 045006 (2016).






























\end{thebibliography}

\end{document}